\documentclass[nologo,11pt,a4paper]{ETHpaper}
\usepackage[english]{babel}
\usepackage{tabularx}
\usepackage{MnSymbol}
\usepackage{booktabs}
\usepackage{caption}
\usepackage{hhline}
\usepackage[table,dvipsnames]{xcolor}
\usepackage{graphicx}
\usepackage[square,numbers,sort&compress]{natbib}
\usepackage{setspace}

\usepackage{appendix}
\usepackage{amsmath,mathtools}
\usepackage[utf8]{inputenc}
\usepackage{xcolor} \usepackage{soul}
\usepackage{hyperref}
\usepackage{xparse,tikz,calc}
\usepackage{pifont}
\usepackage[subrefformat=parens,labelformat=simple]{subfig}

\usepackage{cleveref}
\usepackage{fontawesome5}
\usepackage{graphicx} % DO NOT CHANGE THIS
\usepackage{amsmath}
\usepackage{graphicx}
\usepackage{xcolor}
\usepackage{dcolumn, booktabs}
\newcolumntype{d}[1]{D{.}{.}{#1}}
\begin{document}

\title{Understanding Online Migration Decisions Following the Banning of Radical Communities}
\titlealternative{Understanding Online Migration} 
\renewcommand*{\thefootnote}{\fnsymbol{footnote}}
\author{Giuseppe Russo$^{1}$\footnote{Corresponding author, \texttt{russog@ethz.ch}}, Manoel Horta Ribeiro$^{2}$, Giona Casiraghi$^{1}$, Luca Verginer$^{1}$}
\authoralternative{G. Russo, M. Horta Ribeiro, G. Casiraghi, L. Verginer}
\address{$^{1}$Chair of Systems Design, ETH Zurich, Switzerland\\
  $^{2}$Data Science Laboratory, EPFL, Switzerland}
%  \url{www.sg.ethz.ch}}
\www{\url{http://www.sg.ethz.ch}}

\reference{%\textbf{NOT FOR DISTRIBUTION!}
  (Submitted for publication)} 
\makeframing

\maketitle
\renewcommand*{\thefootnote}{\arabic{footnote}}
\begin{abstract}
The proliferation of radical online communities and their violent offshoots has sparked great societal concern. 
However, the current practice of banning such communities from mainstream platforms has unintended consequences: 
(i)~the further radicalization of their members in fringe platforms where they migrate; and
(ii)~the spillover of harmful content from fringe back onto mainstream platforms.
Here, in a large observational study on two banned subreddits, \texttt{r/The\_Donald} and \texttt{r/fatpeoplehate}, we examine how factors associated with the RECRO radicalization framework relate to users' migration decisions.
Specifically, we quantify how these factors affect users' decisions to post on fringe platforms and, for those who do, whether they continue posting on the mainstream platform.
Our results show that individual-level factors, those relating to the behavior of users, are associated with the decision to post on the fringe platform.
Whereas social-level factors, users' connection with the radical community, only affect the propensity to be coactive on both platforms. 
Overall, our findings pave the way for evidence-based moderation policies, as the decisions to migrate and remain coactive amplify unintended consequences of community bans.
\end{abstract}

\begin{figure}[ht]
\centering
 \includegraphics[width=.825\textwidth]{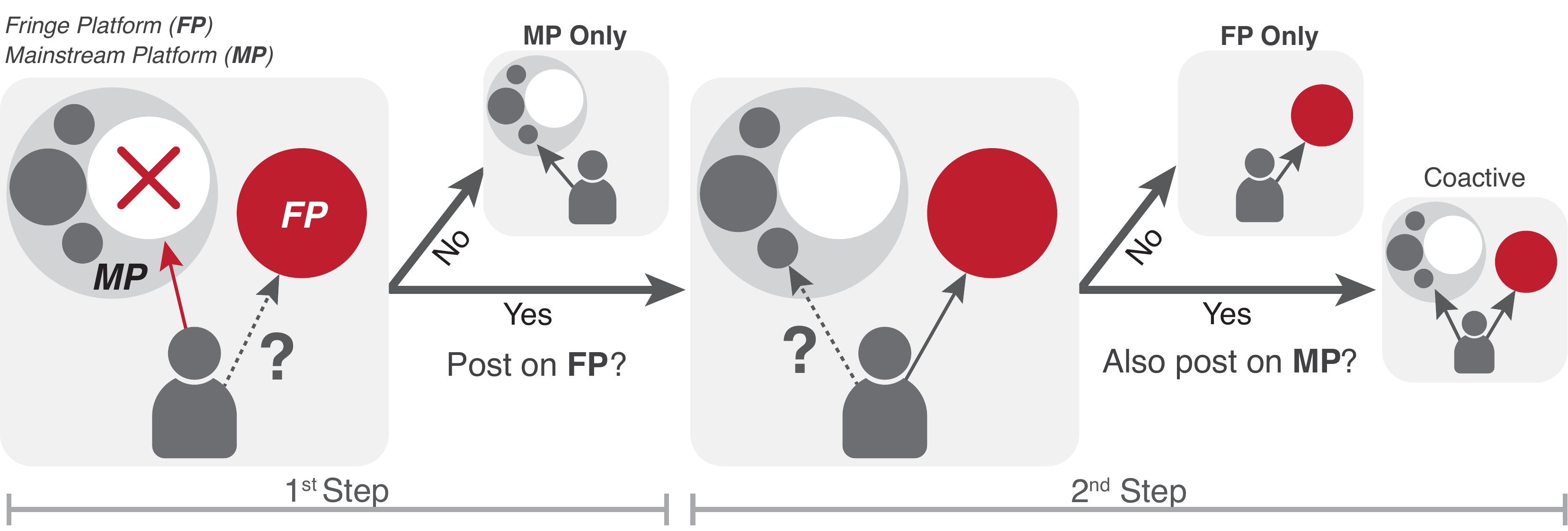}
 \caption{When a community is banned from a mainstream platform for harmful conduct (indicated by an $\times$ in the figure), users have to decide whether to (i) migrate to fringe platforms where the community migrated toward; and (ii) remain active in other communities in the mainstream platform. In this paper, we study factors associated with these two decisions.}
 \label{fig:teaser}
\end{figure}

\section{Introduction}\label{sec:intro}

Online platforms enforce strict moderation policies to prevent the spread of content deemed harmful~\cite{reddit2020, transparency_yt}.
Consequently, they ban communities breaching their guidelines~\cite{qanon2020}, often resulting in the migration of these communities to fringe, alternative platforms with little to no content moderation~\cite{freelon2020false}.

Previous work suggests that the migration of banned communities radicalizes their members~\cite{zuckerman2021deplatforming, ali2021understanding}.
For example, after the prominent Reddit community \texttt{r/The\_Donald} migrated to a self-hosted website, \emph{thedonald.win}, its users became significantly more toxic~\cite{horta2021platform}.
Further, toxic behavior accepted on fringe platforms spills back onto the mainstream platforms through coactive users, i.e., users that remain active on both~\cite{russo2022spillover}.
For example, users migrating to \emph{thedonald.win} \textit{and} continuing to post on Reddit became more toxic also on the latter.
Therefore, users of banned communities decide (i) whether to participate on the fringe platform (\emph{migration decision}) and (ii) whether to participate in both platforms or abandon the original one (\emph{coactivity decision}).
Understanding the factors driving these decisions informs stakeholders of the externalities of community bans, paving the way for more evidence-based moderation policies.

To identify these factors, we build on RECRO, a theoretical model for internet-mediated radicalization proposed by~\citet{neo2019internet}.
RECRO defines five phases of radicalization: Reflection, Exploration, Connection, Resolution, and Operational.
Only the first three phases describe online activity.
The latter two describe real-world actions that are not observed in online data.
For this reason, we focus on studying factors related to Reflection, Exploration, and Connection.
\textit{Reflection factors} describe needs and vulnerabilities that render individuals more receptive to alternate belief systems.
\textit{Exploration factors} quantify how individuals make sense of information put forth by the radical community.
Finally, \textit{Connection factors} describe the influence of community members on each other.
We hypothesize that factors associated with how users reflect, explore, and connect (REC) with radical communities explain their migration decisions following bans.

Our analysis confirms that REC factors are indicative of migration and coactivity decisions.
Interestingly, the factors associated with each decision differ.
While reflection factors correlate mostly with the decision to migrate to the fringe platform, Connection factors are mainly associated with the coactivity decision.
Since Reflection describes an individual's online behavior while Connection describes an individual's social environment, we conclude that individual motives drive the decision to engage with the new platform, whereas social factors drive coactivity.

\section{Related Work}\label{sec:rel_work}

\paragraph{Antisocial online communities}
Various online communities consistently engage in antisocial behavior~\cite{marwick2018drinking}, often harassing minorities and sympathizing with extremist ideologies~\cite{rieger2021assessing}.
These communities have disproportionate influence over memes and news shared on the web~\cite{zannettou2018origins}.
Further, they have been closely associated with medical misinformation~\cite{zeng2021conceptualizing}, conspiracy theories~\cite{sipka2022comparing}, and extremist ideologies~\cite{mcilroy2019welcome}.

Among those communities on Reddit, we consider \texttt{r/The\_Donald} and \texttt{r/fatpeoplehate}.
The subreddit \texttt{r/The\_Donald} was created in June 2015 to support the then-presidential candidate Donald Trump's bid for the U.S. Presidential election.
This community has been closely linked with the rise of the ``alt-right'' movement hosting racist, sexist and islamophobic discussions~\cite{lyons2017ctrl}, and spreading conspiracy theories~\cite{paudel2021soros}.
\citet{flores2018mobilizing} have studied how active participants in \texttt{r/The\_Donald} mobilized the community to engage in ``political trolling.''
The subreddit \texttt{r/fatpeoplehate} was created in June 2015 to promote collective actions of body shaming and violence against overweight people.
In 2015, Reddit banned \texttt{r/fatpeoplehate} following a newly-introduced policy to ban subreddits targeting and harassing specific groups~\cite{chandrasekharan2017you}.

\paragraph{Online Migration}
Both \texttt{r/The\_Donald} and \texttt{r/fatpeoplehate} have been ``de-platformed,'' i.e., banned from Reddit for breaching their guidelines.
In response to the banning, users of these communities migrated to fringe platforms where they continued their discussions.
For example, users of \texttt{r/The\_Donald} migrated \emph{en masse} to \emph{thedonald.win}, a self-hosted website where users of \texttt{r/The\_Donald} could continue discussing and behaving as before.
Similarly, users of \texttt{r/fatpeoplehate} migrated to the Reddit-like platform, \emph{voat.co}, where they re-established their community.

Previous work has studied the effects of deplatforming, finding that after a ban, users reduce their activity on mainstream platforms~\cite{jhaver2021evaluating}, but also that users often migrate to other fringe platforms, where they become more toxic~\cite{ali2021understanding}.
These studies imply that user migration, as an outcome of community deplatforming, might isolate users and expose them to more extreme content.
Additionally, \citet{russo2022spillover} identified a `radicalization spillover' from fringe to mainstream platforms caused by users that migrated to the fringe but remained active on the mainstream platform (referred to as \emph{coactive} users).
Therefore, previous research concludes that online migration of radical communities yields consequences at community and platform levels.
At the community level, migration to fringe platforms can push users to further radicalize and participate in real-world violent events.
At the platform level, users that have decided to migrate but remain active on the mainstream platform, \emph{coactive} users, can undermine the efficacy of moderation policies such as banning.

\paragraph{Online Radicalization} Radicalization is defined as the adoption of extreme political, social, or religious ideals and aspirations that reject or undermine the status quo of society (e.g., acceptance of differences), which can lead to violence to achieve these goals \cite{dalgaard2010violent}.
Multiple radicalization models have been proposed \cite{laquer1998origins, kruglanski2014psychology}.
However, they either ignore internet-mediated radicalization or focus on psychological predispositions \cite{neo2019internet}.
The RECRO model proposed by Neo is a theoretical model for internet-mediated radicalization.
It has been used to study anti-vaccination discussions~\cite{van2019echo} and engagement with conspiracy theories~\cite{phadke2022pathways}.
The RECRO model consists of five phases: Reflection, Exploration, Connection, Resolution, and Operational (RECRO).
As noted by \citet{neo2019internet}, these phases may overlap and occur multiple times in the radicalization process.
Therefore, we consider these phases simultaneously to include such an overlap feature of the RECRO model.

Similarly to \citet{phadke2022pathways}, we focus on the first three phases---Reflection, Exploration, and Connection---based on how individuals engage with radical online content.
The \textbf{R}eflection phase describes an individual's emotional state, making them susceptible to radical narratives.
\citet{neo2019internet} qualitatively find that heightened emotions like aggressiveness and anxiety are linked to \emph{reflection}.
The \textbf{E}xploration phase describes how individuals consume content, look for new information, and get exposed to radical narratives.
Finally, \textbf{C}onnection describes how interactions with members of radical communities influence the individual.

\paragraph{Relation to prior work}
Previous work has found that community-level bans can have negative externalities: users can migrate to other, more toxic communities~\cite{horta2021platform} and may cause `radicalization spillovers' from fringe to mainstream platforms~\cite{russo2022spillover}.
As these consequences arise mainly from user decisions to migrate and to remain coactive, we build upon the RECRO radicalization framework~\cite{neo2019internet} to investigate the factors associated with these decisions.

\section{Materials and Methods}\label{sec:methods}
\subsection{Data}

To study migration decisions, we use data from two subreddits (\texttt{r/The\_Donald} and \texttt{r/fatpeoplehate}; see Section~\ref{sec:rel_work}) and the fringe platforms their users migrated \textit{en masse} after they were banned (\emph{thedonald.win} and \emph{voat.co})
We collect the entire posting history relevant to the two communities on Reddit and the fringe platforms.

\paragraph{Reddit}
Using the Push API~\cite{baumgartner2020pushshift}, we collect the posts made on the two subreddits, starting six months before they were banned.
Specifically, for \texttt{r/fatpeoplehate}, we collect from February 1, 2015, to August 1, 2015; for \texttt{r/The\_Donald}, from November 11, 2019, to February 26, 2020.
For each subreddit, we also collect all contributing users' entire Reddit posting history.

\paragraph{Fringe Platforms}
We obtain \emph{thedonald.win} data using custom Web crawlers and \emph{voat.co} data from Mekacher and Papasavva~\cite{mekacher2022can}.
For each platform, we collect posts made in the 36 weeks around the ban.

We discard users with low activity to ensure  that our analysis is not biased by users with low engagement within the analyzed communities (similarly to~\cite{samory2018conspiracies}).
Precisely, we discard users who made less than ten posts on \texttt{r/The\_Donald} or \texttt{r/fatpeoplehate} \emph{before} the ban.
Further, among those users that post on reddit after the ban we discard those that contributed with less that ten posts on the whole Reddit \emph{after} the ban.
Finally, we assume users were active on the fringe platforms after the ban only if they made at least five posts.
Overall, we collect $2.5$ million posts from \texttt{r/The\_Donald} and \texttt{r/fatpeoplehate}.
These, combined with the data obtained from the other $4,786$ subreddits, yield a total of $91.2$ million posts by  $\sim140,000$ users ($91,244$ for \texttt{r/The\_Donald}, $49,765$ for \texttt{r/fatpeoplehate}).
Moreover, we collect over $2.5$ million posts by $38,510$ users from \emph{thedonald.win}, and $1.3$ million posts from $26,223$ users from \emph{voat.co}.

\newcommand{\mpo}{\text{\textsc{RO}}}
\newcommand{\coa}{\text{\textsc{CA}}}
\newcommand{\fmg}{\text{\textsc{FM}}}
\newcommand{\mg}{\text{\textsc{MG}}}

\begin{figure}[hbt]
\centering
  \includegraphics[width=.3\columnwidth]{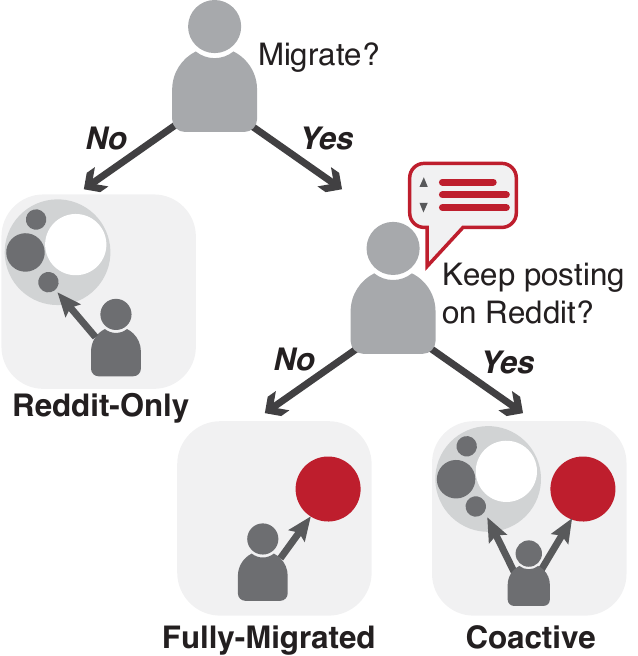}
  \caption{Mapping migration decisions to users' labels.}\label{fig:labels}
\end{figure}

\paragraph{Users labeling}
After the ban, users decide whether to post on the fringe platform or not.
We label users that post on the fringe platform as \emph{migrated} ($\mg$).
Those users that keep posting on Reddit but never post on the fringe platform are labeled as \emph{Reddit-Only} ($\mpo$).
In the second step of the migration, users that posted on the fringe platform may decide if posting on both platforms or to post exclusively on the fringe.
We label those who continue posting on Reddit and the fringe platform as \emph{coactive} ($\coa$).
Differently, users who stop posting on Reddit after the ban and post \emph{exclusively} on the fringe platform are labeled as \emph{Fully-Migrated} ($\fmg$).
To track users across platforms, we apply exact string matching on their usernames, following previous works~\cite{horta2021platform,newell2016user}.
This strategy emphasizes precision over recall: users who have changed their usernames during the migration are not considered migrated.
We account for this fact while interpreting our results.
Importantly, \texttt{r/The\_Donald} had a system to facilitate username continuity across platforms~\cite{donaldpost1}.
In \cref{fig:labels}, we show how these labels are mapped to the different migration decisions.

\subsection{Modelling Migration}\label{sec:dem}

We investigate how radicalization factors drive migration decisions following community banning.
To answer this question, we formalize platform migration as a two-step decision process (see \cref{fig:teaser}).
In the first step, users become active on the fringe platform and start posting there.
In the second step, users posting on the fringe platform choose whether to post on both platforms or to cease all activity on the mainstream platform. 
Previous work suggests that users explore fringe platforms and decide whether to participate based on a variety of reasons~\cite{newell2016user}, e.g., the abundance of niche content in the mainstream platform and the permissive moderation style of fringe platforms.

We model the migration process with a Heckman two-stage regression~\citep{Heckman1979}.
The first stage models the propensity $s_i$ of the user $i$ to post on the fringe platform after the ban (first migration decision). 
The second stage models the likelihood that the user $i$ remains coactive, accounting for their propensity $s_i$ to post on the fringe platform (coactivity decision).
For both stages, for a given user $i$, we consider various factors related to Reflection, Exploration, and Connection referred to as vectors $\mathbf{R_i}$, $\mathbf{E_i}$, and $\mathbf{C_i}$ (see \cref{sec:recro}).
Formally,
\begin{align}
\small
    s_i &\sim \gamma_0+\overbrace{\mathbf{\gamma_1} \mathbf{R_i}+\mathbf{\gamma_2}\mathbf{E_i}+\mathbf{\gamma_3}\mathbf{C_i}}^\text{REC variables} + \gamma_4\text{ER}_i \label{eq:selection_eq} \\
    \coa_i &\sim \beta_0+\underbrace{\mathbf{\beta_1} \mathbf{R_i}+\mathbf{\beta_2}\mathbf{E_i}+\mathbf{\beta_3}\mathbf{C_i} }_\text{REC variables} + \psi(s_i) \label{eq:outcome_eq}\,,
\end{align}
where \eqref{eq:outcome_eq} is the main equation and \eqref{eq:selection_eq} is the selection equation. 
Note that $\psi(s_i)$ accounts for the propensity $s_i$ of user $i$ to post on the fringe platform.
Also, to improve the error estimates, we include an exclusion restriction, as suggested by \citet{puhani2000heckman}.
The exclusion restriction is a variable affecting the selection propensity ($s$) but not the main outcome ($\coa_i$).
Precisely this is the variable $\text{ER}$ in \eqref{eq:selection_eq}.
In our case, the variable must affect the decision to migrate but only weakly affect coactivity.
To this end, we choose \emph{language coherence} (see \cref{sec:recro}), measured as the ability to conform to the language used by the focal community.
Continuing to use the language of the focal community may drive users to follow the group on the fringe platform.
However, language usage does not preclude continued activity on the mainstream platform.

Importantly, if we were to model these stages without the Heckman correction, i.e., considering all users in the first stage and only migrating users in a second, separate regression, we would estimate the regression parameters incorrectly.
Moreover, we would \emph{not} answer the general question: \textit{what is the probability of any user becoming coactive if their community was banned?}
Instead, we would estimate which factors increase the probability that a migrated user ($\mg$) becomes coactive ($\coa$).
The reason for this is selection bias.
Indeed, we only observe coactivity for a subset of the population and not a random and representative sample of the banned community (see \cref{fig:labels}).

\section{Operationalizing RECRO}\label{sec:recro}

We operationalize the  \textbf{R}eflection, \textbf{E}xploration, and \textbf{C}onnection stages of the RECRO framework using language, activity, and interaction features.
These are calculated on the \emph{pre-banning} activities on Reddit of users of \texttt{r/The\_Donald} and \texttt{r/fatpeoplehate}.

\vspace{-3mm}
\subsection{Operationalizing Reflection}
\citet{neo2019internet} describes \textbf{R}eflection as the emotional state making individuals receptive to radical narratives.
To operationalize it,  we compute a set of features quantifying the usage of toxic language, emotionality, anger, and anxiety in users' posts.

\newcommand{\tox}{(TOX)}
\paragraph{Language Toxicity \tox}
Users of radical communities are prone to use toxic language and engage in antisocial behavior, such as harassment, trolling, and cyberbullying~\cite{zannettou2020measuring}.
Following \citet{grover2019detecting} suggestion that antisocial behavior can be captured through automated text analysis, we use the Perspective API~\citep{perspective} to measure language toxicity.
We compute a user's toxicity \tox{} as the mean toxicity score of all their posts written before the ban on Reddit.
Formally, the toxicity of a user $i$ is $\text{\tox}_i = 1/|P_{i}|\sum_{p \in P_{i}} \text{t}(c)$, where $P_{i}$ is the set of posts of user $i$ and $\text{t}(\cdot)$ is the toxicity score of a post.

\newcommand{\emo}{(EMO)}
\paragraph{Emotionality \emo}
We measure the intensity of emotions a user expresses as the average VADER score of their posts before the ban.
VADER is a lexical rule-based sentiment analysis toolkit designed to identify sentiment in social media.
It assigns a score to each word based on its positivity (negativity), and emotional charge.
For example, the word ``good'' is less emotional than ``awesome'' even though both are positive.
Formally, we define $\text{\emo}_i$ of a user $i$ as $1/|P_{i}|\sum_{p \in P_i} \text{VADER}(p)$.
Where $P_{i}$ is the set of posts of user $i$ and $\text{VADER}(\cdot)$ is the VADER score of a post.

\newcommand{\ang}{(ANG)}
\newcommand{\anx}{(ANX)}
\paragraph{Anger \ang{} and Anxiety \anx}
Users of extreme online groups express anger and anxiety in their posts~\cite{wollebaek2019anger}.
We rely on the \emph{Linguistic Inquiry and Word Count} (LIWC) dictionary, counting the proportion of anger \ang and anxiety \anx words a user $i$ uses in their post written before the ban on Reddit.

\subsection{Operationalizing Exploration}\label{subsec:recro_div}
\citet{neo2019internet} defines the \textbf{E}xploration as the period in which individuals make sense of the online information put forth by radical communities.
We operationalize it by measuring (i) the diversity of interests and (ii) the engagement towards subreddits hosting discussions similar to those of \texttt{r/The\_Donald} and \texttt{r/fatpeoplehate}.

\newcommand{\dvr}{(DIV)}
\paragraph{Diversity of interests \dvr}
To capture how frequently users of radical communities interact with other subreddits, we obtain the frequency of posts across subreddits and quantify how diverse user activity is with the \emph{Gini coefficient}.
The Gini coefficient measures the inequality among values of a frequency distribution.
Thus, users contributing to a few subreddits have a low Gini coefficient, while those contributing to many have a higher score.

\newcommand{\eng}{(ENG)}
\paragraph{Engagement with radical communities \eng}
As measured by \dvr, users of radical communities can post in multiple subreddits.
However, they may gravitate towards subreddits similar to their radical community, engaging with it outside a specific subreddit.
To measure this engagement, we compute what proportion of the user’s pre-ban activity is dedicated to posting in subreddits similar to the two focal subreddits (i.e., \texttt{r/The\_Donald} and \texttt{r/fatpeoplehate}).
We refer to this measure as \eng, and we formalize it as $ \text{\eng}_i=\big (\sum_{s_{ij} \in S \setminus s_b} n_j \text{sim}(s_b, s_j)\big)/\big |P_i\big |$
where $S$ is the set of all subreddits, $P_i$ is the set of all posts made by users $i$,  $n_j$ is the number of posts made by user $i$ on the $j$-th subreddit $s_j$, and $s_b$.
$\text{sim}(s_b, s_j)$ is the similarity between the banned subreddit $s_b$ (i.e., \texttt{r/The\_Donald} and \texttt{r/fatpeoplehate}) and subreddit $s_j$ (see \cref{sec:appendix}.
A high \eng{} score indicates that the user contributes primarily to subreddits close to their radical community.

\subsection{Operationalizing Connection}

The last factor we consider in users' radicalization is their \textbf{C}onnection to the radical online community, i.e.,  their influence on the user.
We operationalize the concept by characterizing interactions with other community members.

\newcommand{\dvi}{(DVI)}
\paragraph{Diversity of Interactions \dvi}
Interactions with users exhibiting an \emph{exclusive} interest in the radical communities might increase their sense of belonging to it~\cite{reedy2013terrorism}.
We operationalize the magnitude of these interactions on a user $i$ as the average  Gini coefficient of users $j$ they directly replied to, weighted by the number of comments exchanged between users $i$ and $j$.
Formally, we define diversity of interactions \dvi{} as $\sum_{j \in \mathcal{N}_i} w_{ij} \text{\dvr}(j)/|\mathcal{N}_i|$.
Where $\mathcal{N}_i$ is the set of those users that directly replied to user $i$, $w_{ij}$ is the number of replies between user $j$ and $i$. 
Finally, $\text{\dvr}(j)$ is computed as described in \cref{subsec:recro_div}

\newcommand{\sen}{(SEN)}
\paragraph{Influence of Seniority \sen}
Senior members of radical communities may exert pressure on users to conform to their views.
We proxy the seniority on Reddit using the account's age (the number of days that elapsed from the first post on the community to the ban).
We define \sen{} as the weighted average of the age differences between user $i$ and the other users they directly replied to, i.e.,  $\frac{1}{|N_i|}\sum_{n \in N_i} w_{ij}[\alpha(i)$ $-\alpha(n)]$.
Where $\mathcal{N}_i$ is the set of those users that directly replied to user $i$,  $w_{ij}$ is the number of replies between user $j$ and $i$.
Finally, $\alpha(\cdot)$ is a function returning a user account's age.

\newcommand{\apb}{(APB)}
\newcommand{\ppb}{(PPB)}
\paragraph{Interactions with pre-ban migrated users}
Interactions with other users who have joined the fringe platform might be before the ban (pre-ban migrated users) may create cross-platform ties that, following the ban, increase the odds of migrating to the fringe platform.
We characterize these interactions in two ways:

\noindent
\emph{Active Interaction with pre-ban migrated users \textbf{\apb}:} 
Active interactions are pairwise interactions between users of the radical community not yet posting on the fringe platform, with users already posting there and vice-versa.
This exchange might be a user's first connection with the fringe platform through users already posting there.
Therefore, we count the proportion of such dyadic interactions with pre-ban migrated users normalized by the user's dyadic interactions on Reddit before the ban.

\noindent
\emph{Passive Interaction with pre-ban migrated users \textbf{\ppb}:}
Users not yet posting on the fringe platform may post on threads where already migrated users also post.
If a user posts on such a thread, we say that they interact \emph{passively}.
By posting on the same thread, users may passively consume the content written by pre-ban migrated users without directly engaging with them.
Specifically, we calculate the user’s co-presence with pre-ban migrated users in threads by counting the threads with posts by pre-ban migrated users.
We normalize this measure by the total number of threads the user participates in the pre-ban period, i.e., before the ban.

\subsection{Other Variables}

\paragraph{Language Coherence}
Language Coherence is a concept proposed by \citet{phadke2022pathways,crossett2010radicalization} expressing how well users align with the language of their community.
Especially in discussions in radical communities, members might express their belonging to the community via their language~\cite{koehler2014radical,scrivens2018searching}.
We measure language coherence using a language model capturing the linguistic state of the community.
In particular, we fine-tune BERT on a dataset of posts written before the ban (we do not use these posts to measure language coherence).
Then, given all pre-ban user posts on Reddit $P_i$, we estimate how unexpected the text is according to the language model fine-tuned on the community language.
Specifically, we compute language coherence as the cross-entropy of all posts  $P_i$ for each user $i$ given the language model.

\paragraph{Activity and Account Age}  We add extra variables to \cref{eq:outcome_eq} and \cref{eq:selection_eq} to control for self selection.
Indeed, users who post on the fringe platform might be more active on the mainstream platform to begin with.
Similarly, more senior users might be more motivated to post on the fringe platform as they develop a stronger attachment to the community.
Therefore we use as control variables (i) \emph{user activity} measured as the number of posts that the user made in the six months before the banning, and  (ii) \emph{user seniority} measured as the number of days elapsed from the first post on the community to the ban of the focal subreddit.

\section{Results}\label{sec:results}

We show how radicalization factors affect users' migration with the regression analysis introduced in \cref{sec:methods}.
We perform separate analyses for \texttt{r/The\_Donald} and \texttt{r/fatpeoplehate}.
In both cases, we find that:
(i) Reflection-related factors affects the first migration step, i.e., the decision to post on the fringe platform (FP) after the ban and
(ii) Connection-related factors have a prominent role in the second migration step, i.e., the decision to be coactive on both the mainstream platform (MP) and fringe platform (FP).
We conclude that individual factors primarily affect the first migration step, while coactivity is primarily affected by social factors.
\Cref{table:table_donald}  and \cref{table:fph} report the coefficients for the factors operationalizing Reflection, Exploration, and Connection for \texttt{r/The\_Donald} and \texttt{r/fatpeoplehate}, respectively.

\begin{table}\centering
\caption{Regression table for \texttt{r/The\_Donald}. We show the parameters' estimates for the first and second stages of the Heckman regression.}
\label{table:table_donald}
\begin{tabular}{l D{)}{)}{9)3} D{)}{)}{9)3}}
\toprule
 & \multicolumn{1}{l}{1st Step} & \multicolumn{1}{l}{2nd Step} \\
\midrule
                       &\emph{Selection Eq.}& \emph{Outcome Eq.}\\
\midrule
(Intercept)      & -1.02 \; (0.34)^{**}  &  1.56 \; (0.70)^{*}  \\
\textbf{Reflection}             &                       &                       \\
\quad Toxicity \tox         & 0.78 \; (0.15)^{***}  & -0.91 \; (0.37)^{*} \\
\quad Emotionality \emo     & 1.03 \; (0.20)^{***}  &  0.42 \; (0.44)        \\
\quad Anger \ang            & -1.79\; (1.46)        & -0.29 \; (1.03)       \\
\quad Anxiety \anx         & -0.50 \; (1.12)       &  0.57 \; (0.97)       \\
\textbf{Exploration}            &                       &                       \\
\quad Diversification \dvr & -2.93 \; (0.16)^{***}  & 4.72  \; (1.15)^{***} \\
\quad Engagement \eng      & 1.25 \; (0.25)^{***}  &  0.11 \; (0.51)       \\
\textbf{Connection}             &                       &                       \\
\quad Passive Int. \ppb     & 0.23 \; (0.37)        &  1.56 \; (0.72)^{*}    \\
\quad Active Int. \apb     & -0.01 \; (0.32)       &  7.56 \; (0.49)^{***}  \\
\quad Neigh. Seniority \sen &  0.37 \; (0.29)       &  4.22 \; (1.23)^{***}  \\
\quad Neigh. Div. \dvi     &  -0.09 \; (0.33)      &  2.44 \; (0.66)^{***}  \\
\textit{Controls}      &                       &                        \\
\quad Coherence       & -5.22 \; (0.25)^{***} &                         \\
\quad Numb.\ Posts    &   2.39 \; (1.67)      & 0.55 \; (1.32)          \\
\quad Seniority       &   1.54 \; (0.63)^{*}  & 0.54 \; (0.25)^{*}      \\
\midrule
Rho ($\rho$)           &                       &  0.28 \; (0.01)^{***}  \\
Sigma ($\sigma$)       &                       &  0.32 \; (0.26)  \\
AIC                    &8681.89               & 1433.90               \\
Num. obs.              & 12053                 & 2740                \\
\bottomrule
\multicolumn{3}{l}{\scriptsize{$^{***}p<0.001$; $^{**}p<0.01$; $^{*}p<0.05$}}
\end{tabular}
\end{table}
\begin{table}\centering
\caption{Regression table for \texttt{r/fatpeoplehate}. We show the parameters' estimates for the first and second stages of the Heckman regression.}
\label{table:fph}
\begin{tabular}{l D{)}{)}{9)3} D{)}{)}{9)3}}
\toprule
 & \multicolumn{1}{l}{1st Step} & \multicolumn{1}{l}{2nd Step} \\
\midrule
                       &\emph{Selection Eq.}& \emph{Outcome Eq.}\\
\midrule
(Intercept)      & -2.72 \; (0.57)^{**}  &  2.06 \; (1.2)^{*}  \\
\textbf{Reflection}             &                       &                       \\
\quad Toxicity \tox        & 1.12 \; (0.14)^{***}  & -1.06 \; (0.64) \\
\quad Emotionality \emo    & 0.73 \; (0.20)^{**}  &  0.34 \; (0.52)        \\
\quad Anger \ang           & 0.53\; (1.82)        & -0.12 \; (0.41)       \\
\quad Anxiety \anx         & -0.05 \; (1.30)       &  0.41 \; (0.76)       \\
\textbf{Exploration}            &                       &                       \\
\quad Diversification \dvr & -2.43 \; (0.21)^{***}  & 3.52  \; (0.54)^{***} \\
\quad Engagement \eng      & 1.48 \; (0.52)^{**}  &  0.17 \; (0.58)       \\
\textbf{Connection}             &                       &                       \\
\quad Passive Int. \ppb    & 0.36\; (0.23)        &  1.96 \; (0.46)^{*}    \\
\quad Active Int.  \apb    & -0.05 \; (0.41)       &  5.56 \; (0.88)^{***}  \\
\quad Neigh. Seniority \sen &  1.17 \; (1.09)       &  3.27 \; (1.01)^{**}  \\
\quad Neigh. Div. \dvi     &  -0.22 \; (0.19)      &  1.92 \; (0.66)^{**}  \\
\textit{Controls}      &                       &                        \\
\quad Coherence       & -4.87 \; (0.45)^{**}  &      \\
\quad Numb.\ Posts    &  3.21 \; (2.72)       & 1.88 \; (1.52)          \\
\quad Seniority       &  0.07 \; (0.39)   & 0.51 \; (0.27)^{*}      \\
\midrule
Rho ($\rho$)             &                       &  0.36 \; (0.02)^{***}  \\
Sigma ($\sigma$)            &                       &  0.44 \; (0.12)^{*}  \\
AIC                    & 7953.34               & 1681.89               \\
Num. obs.              & 8168                 & 916                 \\
\bottomrule
\multicolumn{3}{l}{\scriptsize{$^{***}p<0.001$; $^{**}p<0.01$; $^{*}p<0.05$}}
\end{tabular}
\end{table}
\begin{figure*}[ht!]
	\centering
    \newcommand{\centered}[1]{\begin{tabular}{@{}l@{}} #1 \end{tabular}}
    \bgroup
	\def\arraystretch{0}
    \begin{tabular}{>{\centering\arraybackslash} m{.3\textwidth} >{\centering\arraybackslash} m{.3\textwidth} >{\centering\arraybackslash} m{.3\textwidth} >{\centering\arraybackslash\hspace{-1em}} m{.05\textwidth}}
        {\phantom{aaaaa}\huge \textbf{R}eflection} & {\phantom{aaaaa}\huge \textbf{E}xploration} & {\phantom{aaaaa}\huge \textbf{C}onnection} &\\
        \subfloat[\label{fig:reflection1}]{%
        \includegraphics[width=0.35\textwidth]{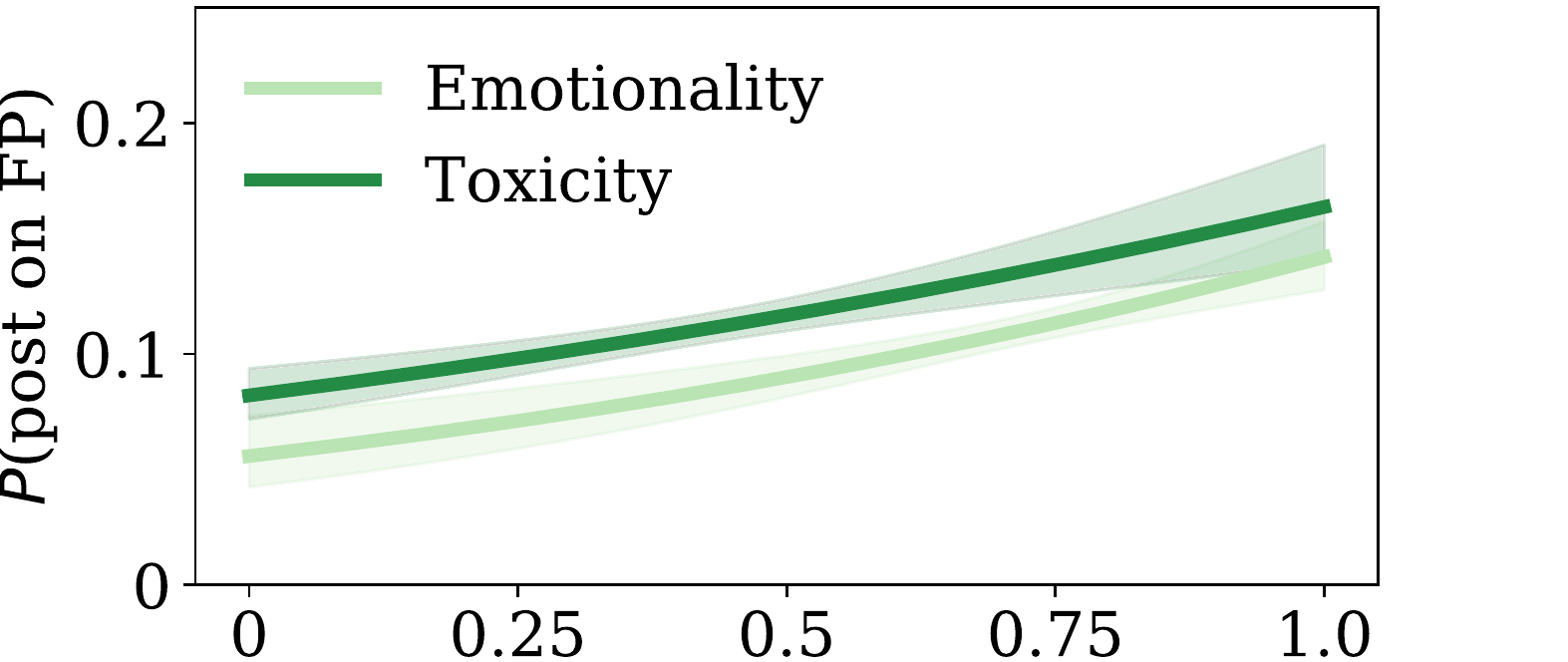}}
        &%\hspace{\fill}
        \subfloat[\label{fig:exploration1} ]{%
        \includegraphics[width=0.35\textwidth]{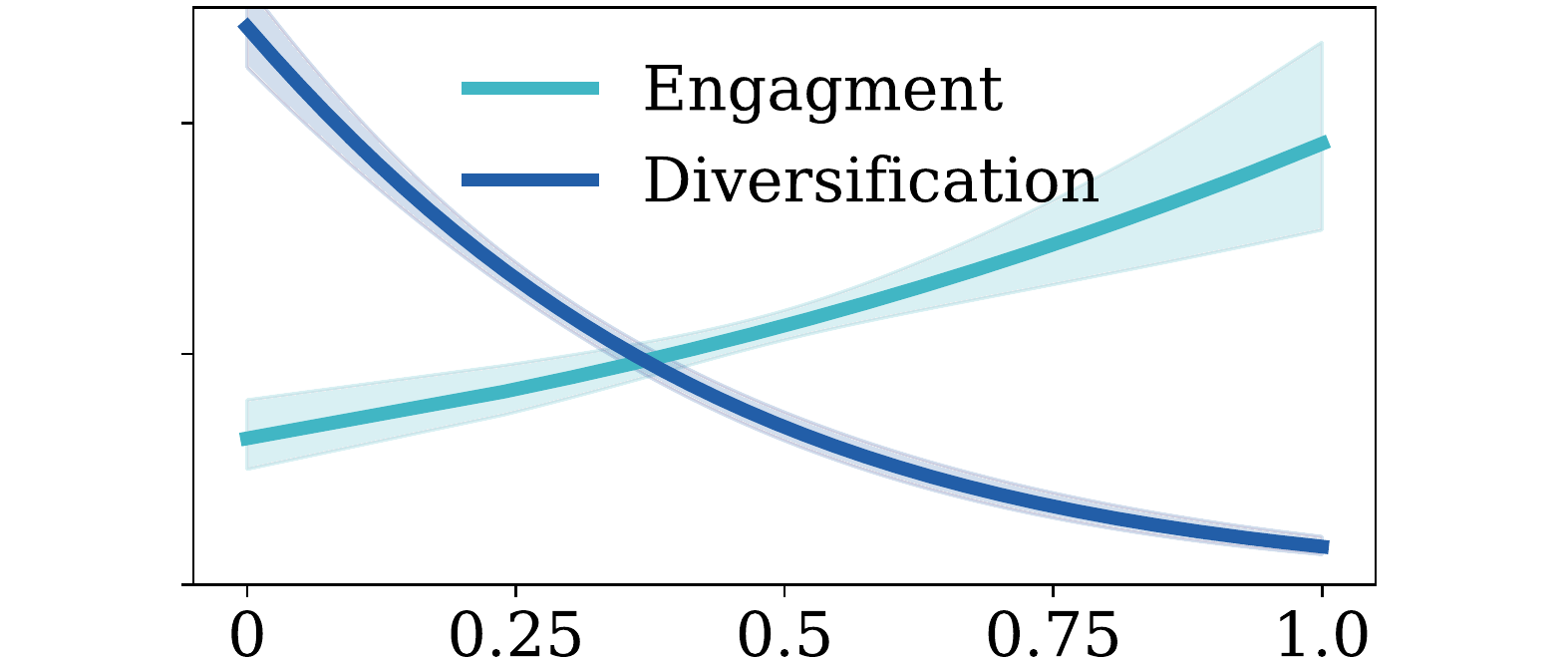}}
        &%\hspace{\fill}
        \subfloat[\label{fig:connection1}]{%
        \includegraphics[width=0.35\textwidth]{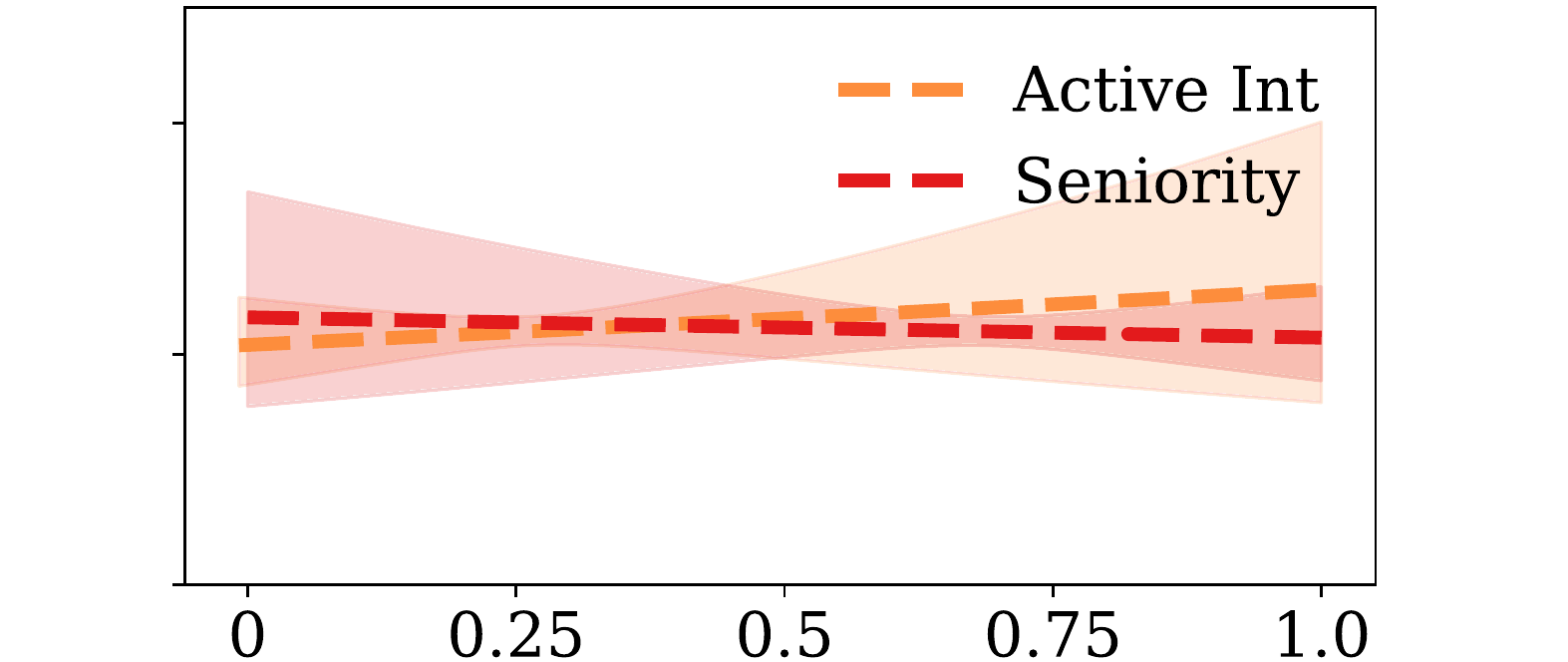}} &
        \centered{\rotatebox[origin=c]{-90}{{\large \textbf{(1st Step)}}}}\\
    \subfloat[\label{fig:reflection2}]{%
    \includegraphics[width=0.35\textwidth]{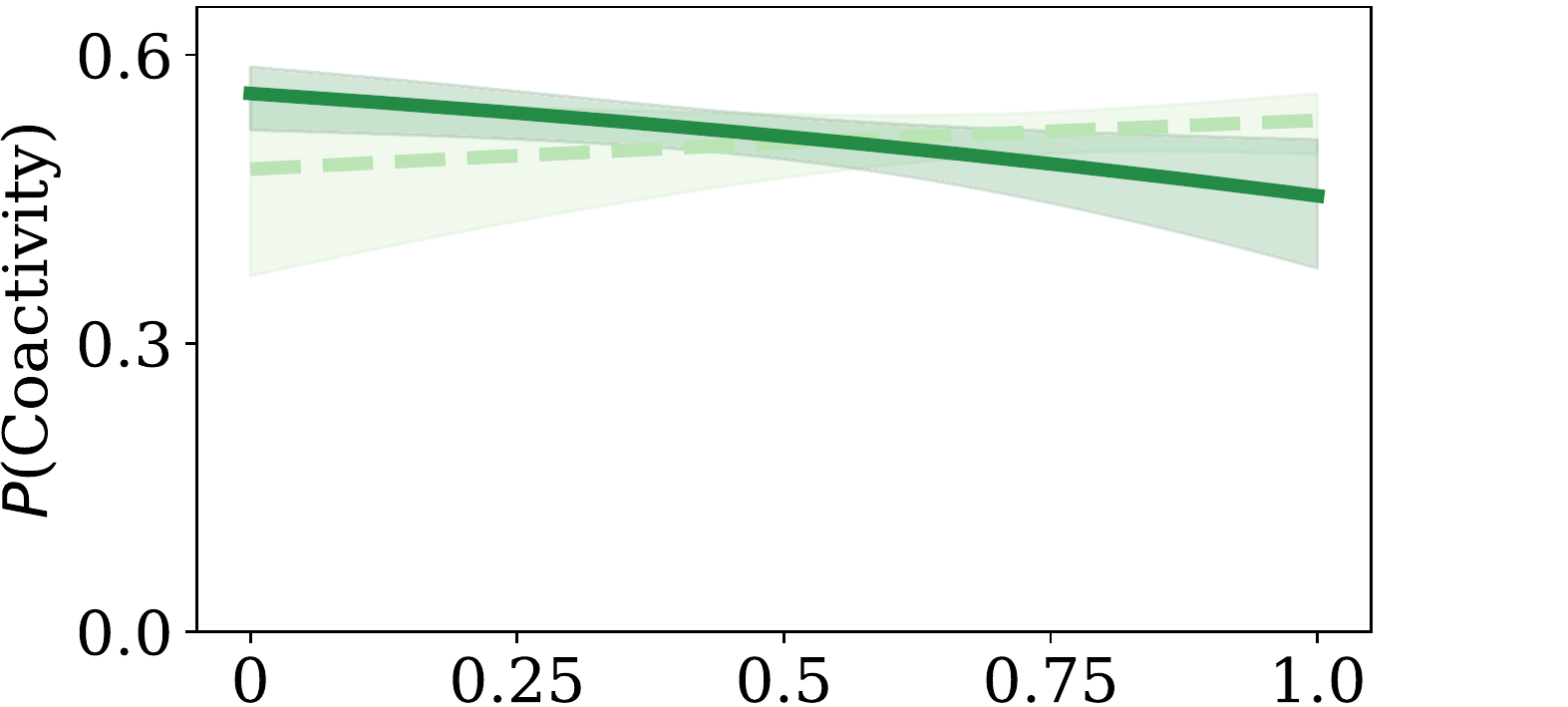}}
    &%\hspace{\fill}
    \subfloat[\label{fig:exploration2} ]{%
    \includegraphics[width=0.35\textwidth]{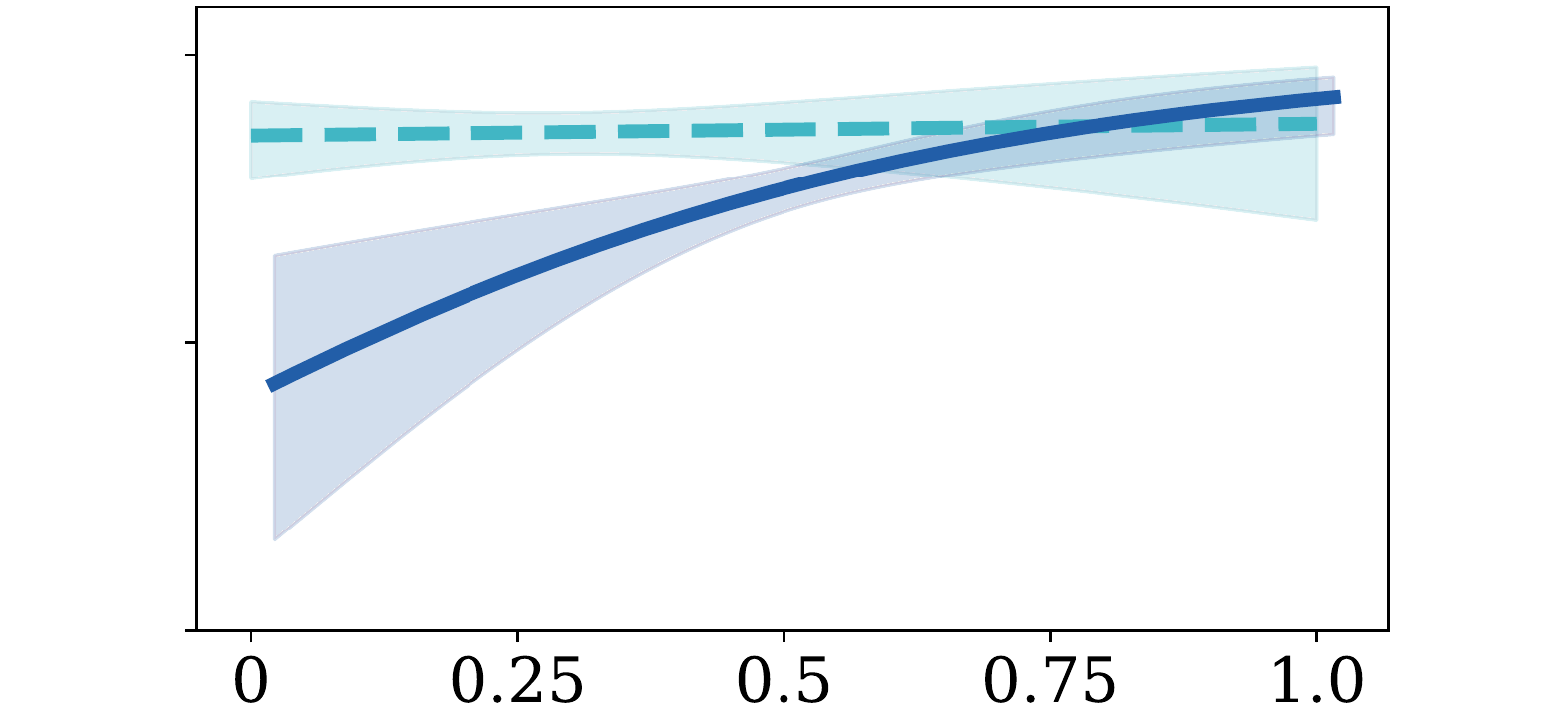}}
    &%\hspace{\fill}
    \subfloat[\label{fig:connection2}]{%
    \includegraphics[width=0.35\textwidth]{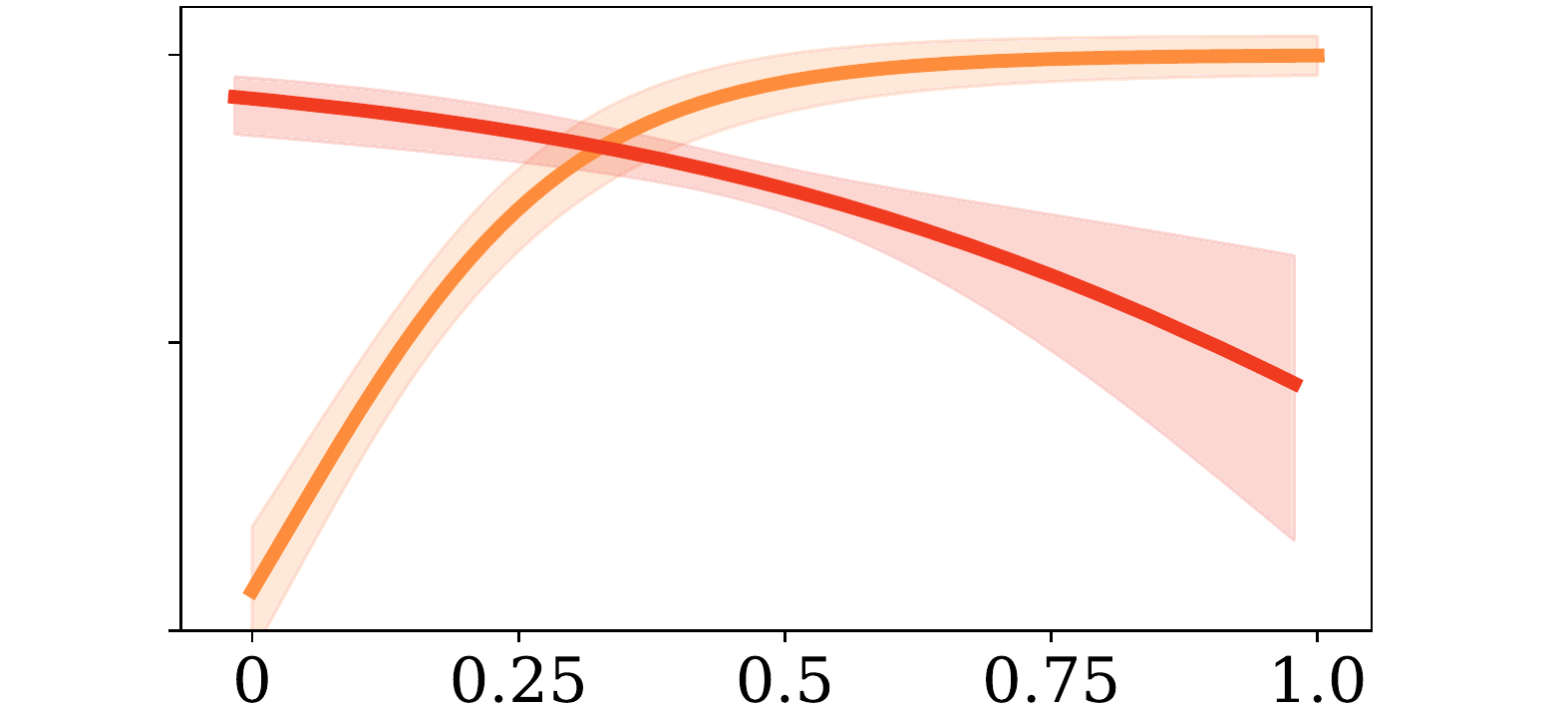}} &
    \centered{\rotatebox[origin=c]{-90}{{\large \textbf{(2nd Step)}}}}
    \end{tabular}
    \egroup
    \caption{Marginal effects of REC factors for \texttt{r/The\_Donald}. Solid lines represent factors with a significant effect. Dashed lines represent factors that do not have a significant effect. The shaded areas show the 95\% confidence intervals. The marginals have been estimated assuming all other parameters are kept constant to their observed average values. \emph{(top row)} Marginal effects for the first migration step. \emph{(bottom row)} Marginal effects for the second migration step.}\label{fig:all}
    \end{figure*}

\subsection{First Step: Migration to Fringe Platforms}

\paragraph{Reflection}
Higher toxicity and emotionality on the MP are associated with more posts on the FP, as both \tox{} and \emo{}  have positive coefficients for \texttt{r/The\_Donald} ($\beta_{\text{TOX}}^{TD}=0.78$, $\beta_{\text{EMO}}^{TD}$=$1.03$), and \texttt{r/fatpeoplehate} ($\beta_{\text{TOX}}^{FPH}$=$1.12$, $\beta_{\text{EMO}}^{FPH}$=$0.73$), respectively.
In \cref{fig:all}, we show the marginal effects on the probability of posting on the FP.
For example, we observe that, for \texttt{r/The\_Donald}, a $20\%$ increase in \tox{} increases the probability of becoming active on the FP by $8\%$ (c.f., \cref{fig:reflection1}).
An analysis of emotionality yields similar results.
Reflection factors describing users' \emph{individual} characteristics are significant predictors of the first migration decision.
Interestingly, we find that the effect of \ang{} and \anx{} is not statistically significant.

\paragraph{Exploration} We find that Exploration factors, engagement \eng{} and diversity \dvr{}, are associated with the first migration decision.
Specifically, the lower users' \dvr{} is ($\beta_{\text{\dvr}}^\text{TD}=-2.93$, $\beta_{\text{\dvr}}^\text{FPH}$=$-2.43$), the more likely they are to post on the FP.
For example, a $30\%$ decrease in \dvr{} increases the probability of posting on the FP by $18\%$ (c.f., \cref{fig:exploration1}).
Since \dvr{} characterizes the heterogeneity of user interests, it acts as a `pull factor' hindering the tendency to post on the FP.
This result is in line with the survey study by \citet{newell2016user}.
Additionally, we find that users participating in other subreddits related to \texttt{r/The\_Donald} or \texttt{r/fatpeoplehate} (\emph{engagement}) are more likely to post on the fringe platform after the ban on the MP.
In synthesis, Exploration, interpreted as the utilization of the mainstream platform, predicts the first migration step well.

\paragraph{Connection.} We find that factors characterizing the connection to other community members are not statistically significant and, thus, do not correlate to the first migration decision. This is also shown in \cref{fig:connection1}, the marginal effects of the features characterizing the social factors do not affect the probability of posting on FP.

\subsection{Second step: Coactivity across Platforms}

\paragraph{Reflection} Reflection factors determine users' coactivity to a minor extent.
Toxicity \tox{} is the only significant reflection factor in the case of \texttt{r/The\_Donald}, users that are less toxic in the mainstream platform have a higher chance of being coactive after the ban ($\beta_{TOX}^{TD}=-0.91$).
For \texttt{r/fatpeoplehate}, no reflection factor is statistically significant.
In \cref{fig:reflection2}, we show that both \tox{} and \emo{} do not increase the probability of being coactive.

\paragraph{Exploration} \emph{Diversification} affect users' coactivity as in the first migration step.
The lower the users' \dvr{}, the more they behave as coactive users ($\beta_{DIV}^{TD}=4.72$, $\beta_{DIV}^{FPH}=3.52$).
For instance, in \cref{fig:exploration2}, we show for \texttt{r/The\_Donald} that a 30\% decrease in \dvr{} increases by 22\% the probability of coactivity.
Unlike the first migration step, \eng{} does not affect coactivity.

\paragraph{Connection} All Connection's factors  become statistically significant and strongly affect users' coactivity.
In particular, from \cref{table:table_donald} and \cref{table:fph}, we observe that the more users directly interact with pre-ban migrated users \apb{}, the less they tend to stay coactive.
Further, we find that users interacting with more recent members are more likely to be active on both platforms after the ban \sen{}.
In \cref{fig:connection2}, we observe that directly interacting with pre-migrated users increases the probability of coactivity by up to $70\%$ in the case of \texttt{r/The\_Donald} subreddit.

\section{Discussion}

This article studies the post-ban migration of users of radical communities from mainstream to fringe platforms.
Migrating users of radical communities, i.e., those who become active on fringe platforms, either abandon the mainstream platform or remain coactive on both.
Previous work has shown that the migration decision following radical community bans has negative externalities.
Namely 
(i) the creation of more toxic communities elsewhere~\cite{horta2021platform} (if many users decide to migrate); and 
(ii) the spillover of toxic behaviour from the fringe onto the mainstream platform (if many users decide to be coactive).

To understand the factors associated with each decision (migration and coactivity), we conduct a two-step regression analysis.
The first step estimates the probability of users posting on the fringe platform (migration decision).
The second step estimates the propensity of users to be coactive (coactivity decision).
Specifically, we examine two subreddits, \texttt{r/The\_Donald} and \texttt{r/fatpeoplehate}, associated with toxic behavior.

Our analysis reveals how factors of the RECRO radicalization framework relate to users' migration decisions.
Our findings are threefold.
First, individuals' online behavior, described by Reflection, capturing the needs and vulnerabilities of users, is linked with the decision to post on the fringe platform. For example, users that exhibit higher toxicity \tox{} or emotionality \emo{} migrated more often.
Second, interactions with the social environment described by Connection relate \emph{primarily} to the decision to be coactive on both platforms. Surprisingly, we find that interactions with pre-ban migrated users \apb{}/\ppb{} increase the propensity to post on the fringe platform. 
Instead, users with more such interactions are more likely to remain coactive.
Third, the diversity of users' interests, captured by Exploration, hinders them from abandoning the mainstream platform altogether and increases their propensity to remain coactive. This is in accordance with previous findings suggesting that the diversity of content on mainstream platforms like Reddit is a pull factor that limits full migration~\cite{newell2016user}.
These findings suggest that the decision to engage with the new platform is linked to individual motives, whereas social factors are associated with coactivity.

\paragraph{Implications} 
Our work has two implications.
First, our analysis sheds light on how migration decisions mediate the benefits of banning.
Thus, paving the way for moderators to take more informed decisions on  banning.
For instance, platforms like Reddit could estimate how users will react before carrying out community bans.
Second, our findings contribute to a growing literature on understanding online radicalization.
Fringe platforms have been tightly linked with terrorist attacks and extremist ideologies~\cite{zeng2021conceptualizing}, and therefore, studying what makes users migrate to fringe platforms advances our understanding of radicalization.

\paragraph{Limitations}
Our classification of users according to their posting activity on fringe platforms may be inaccurate and thus bias our results.
For example, some users may change their username or only read posts on the fringe platform, and we would erroneously classify them as Reddit-only when they are participating or consuming content from the FP.
Our result remains unbiased, however, if there is no systematic difference between users keeping their usernames across platforms and those changing them.

\small \setlength{\bibsep}{1pt}

\appendix

\section{Appendix}\label{sec:appendix}
\subsection{Subreddit Similarity}

We define a similarity between all subreddits and the two focal subreddits (\texttt{r/The\_Donald} and \texttt{r/fat} \texttt{peoplehate}) considering their polar opposites (\texttt{r/HillaryClinton} and \texttt{r/fatlogic}, respectively).
Given a focal subreddit $s_i$, we define a subreddit $s_{j}$ to be \emph{relevant} for $s_i$ if at least ten users of either the focal subreddit $s_i$ or its polar opposite $s_{\hat i}$ posted at least five times on $s_j$ before the ban.
We then construct a weighted graph for a focal subreddit $s_i$ and its polar opposite $s_{\hat i}$ (e.g., \texttt{r/fatpeoplehate}-\texttt{r/fatlogic}).
The nodes of the graphs consist of (i) $s_i$ and $s_{\hat i}$, and (ii) all relevant subreddits for the two polar opposites.
We draw a weighted edge between two nodes if the corresponding subreddits share at least five active users.
The weight corresponds to the number of users shared.
As a next step, we train the Node2Vec~\cite{grover2016node2vec} algorithm on the graphs built for the pairs \texttt{r/The\_Donald}-\texttt{r/HillaryClinton} and \texttt{r/fatpeoplehate}-\texttt{r/fatlogic}.
As a result, we obtain embeddings specifically catered towards finding subreddits hosting discussions similar to the focal subreddits.
We use the cosine similarity to map subreddits' similarity on a scale from -1 to +1, where +1 represents the higher similarity to the focal subreddits.
To validate our similarity scale, we calculate Spearman's rank-order correlation between the 1,000 subreddits most similar to \texttt{r/The\_Donald}, and \texttt{r/fatpeople}\texttt{hate} and the ranking of publicly available subreddit embeddings by \citet{waller2021quantifying}.
The embeddings from \citet{waller2021quantifying} are not explicitly trained towards finding similarities between specific communities, but they provide a general measure of subreddit similarity.
We find a correlation of 0.64 with p < 0.05, indicating that our scale successfully measures similarity to \texttt{r/The\_Donald} and \texttt{r/fatpeople}\texttt{hate}.
We manually inspect the top 50 subreddits on the similarity scale and confirm that they host discussions similar to \texttt{r/The\_Donald} and \texttt{r/fatpeoplehate}, respectively.

\subsection{Predicting Migration Steps}
\begin{table}[!htb]\centering
\caption{F1-Scores for the different classification models. The first column reports the classification scores on the three labels $\mpo$, $\coa$, and $\fmg$. The second and third column report the classification scores of the first ($\mpo\leftrightarrow\mg$) and second step ($\coa\leftrightarrow\fmg$) respectively of the 2STEP classifier.}\label{tab:predictions}
\small
\begin{tabular}{lccc}
\hline
         & \textbf{\begin{tabular}[c]{@{}c@{}}F1-Score \\ (All)\end{tabular}} & \textbf{\begin{tabular}[c]{@{}c@{}}F1-Score \\ (1st step)\end{tabular}} & \textbf{\begin{tabular}[c]{@{}c@{}}F1-Score\\  (2nd Step)\end{tabular}} \\ \hline
SVM                  & 0.44                                                               & $\sim$                                                                  & $\sim$                                                                  \\
Random Forest        & 0.52                                                               & $\sim$                                                                  & $\sim$                                                                  \\
XGB                  & 0.65                                                               & $\sim$                                                                  & $\sim$                                                                  \\
AdaBoost             & 0.41                                                               & $\sim$                                                                  & $\sim$                                                                  \\
\textbf{2STEP (Our)} & \textbf{0.74}                                                      & 0.84                                                                    & 0.88                                                                    \\ \hline
\end{tabular}
\end{table}

In the analysis presented in \cref{sec:methods} and \cref{sec:results}, we show how reflection, exploration, and connection factors affect user migration. 
Here, we investigate if and how well we can predict users' migration decisions in reaction to a ban.
To answer these questions, we define a hierarchical classification model.
This classifier first distinguishes users who post on a fringe platform from those who will not (first decision).
Subsequently, a second classifier predicts if they become co-active or not.
This second classifier is trained with those users that, according to the first classifier, will post on the fringe platform. 
We use for both the first and second classifier a Gradient Boosting Classifier.

To correct the high imbalance that primarily affects the first classifier, we downsample the users that did not post on the fringe platform as they are ten times more common than those that post on the fringe.
We evaluate our hierarchical classifier against other models using multiple metrics testing prediction capacities.
All the experiments have been conducted by running a five-fold cross-validation using a 60\%-20\%-20\% training-validation-test split.

Our results show that our model outperforms other baselines that do not consider the two-step structure of online migration.
Specifically, we compare the performance of our classifier against baselines that directly predict the final user decision in a single step (Reddit-Only, Coactive, or Fully-Migrated).
Unlike our classifier, these baselines predict users' migration decisions in a single step.
In our experiments, the 2STEP classifier outperforms all baselines by a maximum of 12\% using f1-score accuracy.
These results provide solid evidence that our two-step characterization of the migration process is a valid assumption.
We report the results of our classification in \cref{tab:predictions}.
Such results provide additional support to the analysis that we conducted in \cref{sec:dem}

\end{document}